\documentclass[10pt,conference]{IEEEtran} 
\usepackage[utf8]{inputenc}
\usepackage[T1]{fontenc}
\usepackage[x11names,svgnames,dvipsnames,table,rgb]{xcolor}
\usepackage[english]{babel}
\usepackage{amsmath, amsfonts, amssymb}
\usepackage{csquotes}
\usepackage{graphicx}
\usepackage{mathtools,amssymb,amsthm,amsfonts,nicefrac,amsmath,braket,bbm}
\usepackage{physics}
\usepackage{siunitx}
\usepackage{comment}
\usepackage{xcolor}
\usepackage{tabularx,booktabs,multirow}
\usepackage[english]{babel}
\usepackage[nottoc]{tocbibind}

\usepackage{enumitem}
\usepackage{xargs}

\usepackage[hidelinks]{hyperref}

\hypersetup{
    unicode=false,          
    pdftoolbar=false,        
    pdfmenubar=true,        
    pdffitwindow=false,     
    pdfstartview={FitH},    
    pdftitle={Local robust shadows on a trapped ion computer},    
    pdfauthor={Jadwiga Wilkens, Milena Guevara Bertsch, Marwa Marso, Juris Ulmanis, Albert Frisch, Ingo Roth, Richard Kueng},     
    pdfsubject={},   
    pdfcreator={},   
    pdfproducer={}, 
    pdfkeywords={}, 
    pdfnewwindow=true,      
    colorlinks=true,       
    linkcolor=red,          
    citecolor=blue,        
    filecolor=teal,      
    urlcolor=blue           
}

\usepackage{cite}

\addto\captionsenglish{%
}
\addto\extrasenglish{%
}

\addto\captionsenglish{%
}
\addto\extrasenglish{%
}

\addto\captionsenglish{%
  \def\figurename{Fig.\!}%
}
\addto\extrasenglish{%
}

   \makeatletter
    \newcommand{\linebreakand}{%
      \end{@IEEEauthorhalign}
      \hfill\mbox{}\par
      \mbox{}\hfill\begin{@IEEEauthorhalign}
    }
    \makeatother
\makeatletter
\newcommand{\newlineauthors}{%
  \end{@IEEEauthorhalign}\hfill\mbox{}\par
  \mbox{}\hfill\begin{@IEEEauthorhalign}
  }
  
\makeatother

\makeatletter
\def\tagform@#1{\maketag@@@{\ignorespaces#1\unskip\@@italiccorr}}
\let\orgtheequation\theequation
\def\theequation{(\orgtheequation)}
\makeatother
\let\orgautoref\autoref
\renewcommand{\autoref}[1]{\def\equationautorefname{Eq.}\orgautoref{#1}}
\renewcommand{\baselinestretch}{1.0}

\begin{document}

\title{
Local robust  shadows on a trapped ion computer---a case study
}

\author{
\IEEEauthorblockN{
J. Wilkens\IEEEauthorrefmark{1},
M. Guevara Bertsch\IEEEauthorrefmark{2},
M. Marso\IEEEauthorrefmark{1},
M. Zangerl\IEEEauthorrefmark{2},
F. Girtler\IEEEauthorrefmark{2},
A. Frisch\IEEEauthorrefmark{2},
J. Ulmanis\IEEEauthorrefmark{2},
I. Roth\IEEEauthorrefmark{3},
R. Kueng\IEEEauthorrefmark{1}
}

\IEEEauthorblockA{\IEEEauthorrefmark{1}
\textit{Department for Quantum Information and Computation (QUICK)},
\textit{Johannes Kepler University Linz} \\
Linz, Austria \\
\{jadwiga.wilkens, maroua.marso, richard.kueng\}@jku.at \\
}

\IEEEauthorblockA{\IEEEauthorrefmark{2}
\textit{Alpine Quantum Technologies GmbH} \\
Innsbruck, Austria \\
\{milena.guevara-bertsch, mederika.zangerl, florian.girtler, albert.frisch,  juris.ulmanis\}@aqt.eu \\
}

\IEEEauthorblockA{\IEEEauthorrefmark{3}
\textit{Quantum Research Center},
\textit{Technology Innovation Institute} \\
Abu Dhabi, United Arab Emirates \\
ingo.roth@tii.ae \\
}
}

\maketitle

\begin{abstract}
    We experimentally demonstrate local robust shadows on a trapped-ion quantum computing system, a protocol developed to counteract measurement errors. 
    We alternate between a calibration stage and the shadow estimation stage and also introduce Pauli-X-twirling before measurements in both stages to symmetrize error rates.
       We then demonstrate the protocol on a trapped-ion quantum computer with artificially shortened measurement pulse duration. This yields faster experiments at the cost of increased error rates which are subsequently mitigated by the robust shadow protocol.
    We benchmark this approach on three exemplary quantum states: a local Haar random state, as well as standard and Pauli-correlation-encoded QAOA states. 
    In all three cases, the local robust shadow protocol succeeds at mitigating the increased error rates hailing from shorter measurement pulse durations.
 \end{abstract}


\section{Introduction}

The rapid progress of trapped-ion quantum computing systems has enabled the preparation, manipulation, and control of increasingly large quantum systems~\cite{Hainzer_2024, Kiesenhofer_2023, Kranzl_2022, campbell_path_2025, moses_race_2023, aqt_2026}.
However, current and near-term qubits remain susceptible to noise. 
Beyond this fundamental challenge, extracting information from such quantum systems poses an additional challenge: full quantum state tomography~\cite{Banaszek_2013} requires a number of measurements that scales exponentially with the number of qubits, i.e. the system size~\cite{Haah2016,
Kueng2017, Lowe_2025}.
This fundamental bottleneck has motivated the development of measurement-efficient protocols that still allow one to predict many properties of a quantum state from a limited number of samples without requiring full state tomography.
One prominent example is the framework of classical shadows~\cite{
huang_predicting_2020,Elben_2022}
which enable the simultaneous estimation of exponentially many observables from a single experimental data set.
This sequential read-out protocol has become an essential, platform-agnostic tool for extracting information from unknown quantum systems.

In practice however, classical shadow protocols are not robust against measurement errors.
If not properly accounted for, such imperfections in the measurement device lead to systematic biases in the estimated observables. 
This issue has motivated the development of noise-robust variants of shadow estimation.
Initial approaches incorporated error mitigation techniques for diagonal observables~\cite{berg_model-free_2022, Maciejewski2020mitigationofreadout}, while the robust shadow framework~\cite{chen_robust_2021} has enabled unbiased estimation of general observables, provided the noise does not break the symmetry of the measurement ensemble.
Subsequent work further extended this approach, for example by incorporating state symmetries~\cite{Zhao_2024}, reducing circuit depth in ultra-shallow variants~\cite{Farias_2025} or addressing gate-dependent noise \cite{Brieger2023CompressiveGST}.
Experimental demonstrations of robust shadow protocols have been realized on superconducting platforms~\cite{vitale_2024, hu_demonstration_2025}.

On trapped-ion devices, local classical shadows have been demonstrated in~Ref.~\cite{joshi_mpemba_2024}.
Furthermore, Ref.~\cite{joshi_fcs_2025} implements a related strategy by preparing a trusted state to characterize read-out errors.
However, this approach does not employ an explicit Pauli-X-twirling layer prior to measurement, which is a key ingredient in standard robust shadow constructions.

\begin{figure*}[htbp]
    \centering
    \includegraphics[width=0.98\linewidth]{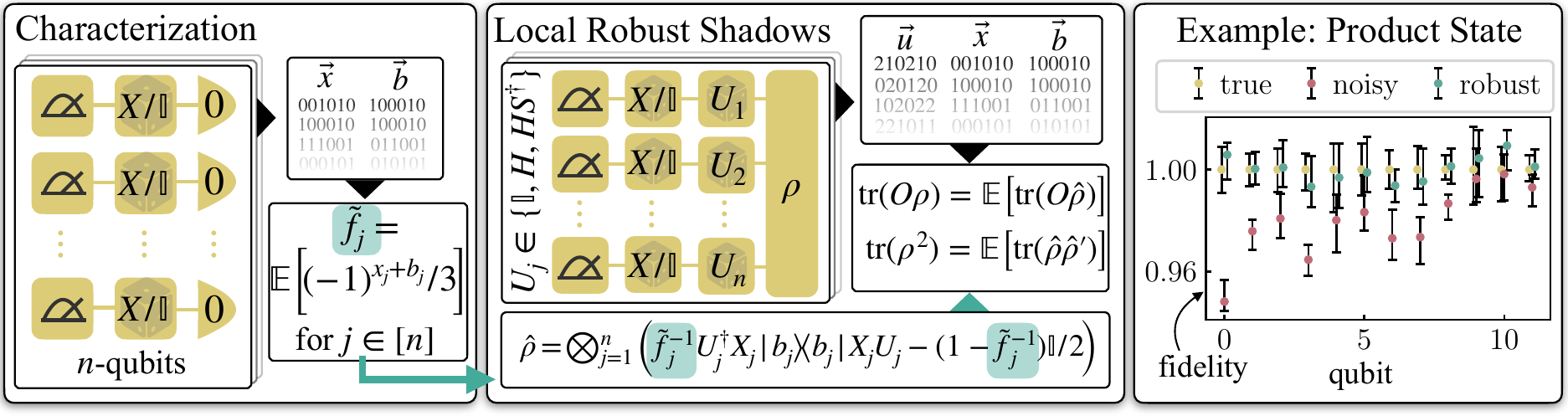}
    \caption{\emph{Local robust shadows with Pauli-$X$-twirling.} Here, circuit diagrams go from left to right. Step I (left): characterize the read-out noise via twirling the zero state with a random bit-flip layer before read-out. 
    Use the acquired data to estimate the noisy single-qubit expansion coefficients $\tilde f_j$.
    Step II (center): execute a sequential local shadow protocol with random single-qubit measurements and add an additional layer of random bit-flips before reading out to symmetrize the read-out error.
    Use the estimated noisy expansion coefficients to create a robust snapshot for the following classical post-processing.
    The plot on the right displays results for estimating the single qubit fidelities of a local Haar-random $12$-qubit state on a trapped-ion quantum computer with an intentionally shortened measurement pulse length of \SI{150}{\micro\second}. This reduces the measurement duration at the cost of increased readout noise which is then mitigated by our protocol.
    }
    \label{fig:main-figure-protocol}
\end{figure*}

In this work, we refine and deploy a completely local, robust classical shadows pipeline on a trapped-ion quantum computing system. To this end, we first establish a direct connection between existing robust shadow protocols~\cite{berg_model-free_2022, chen_robust_2021, Farias_2025}: all these approaches can be unified within a simplified framework based solely on Pauli-X-twirling for the calibration phase and prior to measurement.
This leads to a particularly simple and experimentally efficient implementation that is illustrated in Fig.~\ref{fig:main-figure-protocol}.
Furthermore, we derive explicit expansion coefficients for local noise in terms of the irreducible representations of the measurement channel, together with a simplified estimator tailored to this setting.

Measurements of trapped-ion qubits are extremely accurate but comparatively slow and our framework allows to balance tradeoffs in-between. 
We explore a regime in which the measurement pulse duration is intentionally reduced, thereby increasing read-out errors while decreasing experimental runtime.
We demonstrate that these additional errors can be effectively mitigated within the robust shadows framework, without introducing measurable cross-talk in the read-out process.
Finally, we validate our approach in several representative use cases, demonstrating that local robust classical shadows reliably mitigate measurement errors not only in theory, but also in practice.

\section{Methods}
We first recapitulate the robust shadows framework~\cite{chen_robust_2021}, specialize to local Clifford measurements, and derive our simplified implementation using only X-twirling in Sec.~\ref{sec:results}.

\subsection{Robust shadows}
Classical shadow estimation protocols \cite{hu_demonstration_2025} aim to simultaneously estimate many properties of a $n$-qubit quantum system $\rho$ and 
are based on inverting the average measurement process modeled as a measurement channel $\rho\mapsto\mathcal M(\rho)$ in classical postprocessing.
More concretely, estimating the expectation value of an observable $O$ is related to the measurement channel via the identity $\langle O\rangle_\rho =\Tr(O\rho) = (O|\mathcal M^{-1}\mathcal M |\rho)$.
Here, round brackets denote the Hilbert-Schmidt inner product defined for two linear operators as $(A|B)=\Tr(A^\dagger B)$.

Viewed from this angle, the $n$-qubit computational basis measurement produces a measurement channel $\mathcal M_Z$ consisting of projective $Z$-measurements for each qubit.
Writing $\sigma_0=\mathbb I_2/\sqrt{2}$ and $\sigma_Z=Z/\sqrt{2}$,
this reads as $\mathcal M_Z = \left(|\sigma_0)\!(\sigma_0| + |\sigma_Z)\!(\sigma_Z|\right)^{\otimes n}$
and projects onto the subspace spanned by $\sigma_0$ and $\sigma_Z$ while erasing all components encoded in $X$ and $Y$ directions.
Hence, quantum states cannot be fully recovered once they pass through $\mathcal{M}_Z$.

The computational basis measurement channel can be turned into an invertible map by introducing an extra unitary channel twirling under a suitable group $\mathbb G$ with its representation $U \mapsto \omega(U)(\cdot) = U(\cdot)U^\dagger$:
\begin{align}
    \label{eq:measurement_channel_M_only_twirling}
    \mathcal M = \mathop{\mathbb E}_{U\in\mathbb G}\omega(U)^\dagger\mathcal M_Z\omega(U)\,.
\end{align}

Schur's Lemma implies that such an average decomposes into a linear combination of projectors $\Pi_\lambda$ onto the irreducible subspaces $R_{\mathbb G}$ of the group representation as
\begin{align}
    \label{eq:measurement_channel_M_expansion_coefficients_noiseless}
    \mathcal M =\sum_{\lambda\in\mathrm R_\mathbb G}f_\lambda \Pi_\lambda, && f_\lambda = \Tr(\mathcal M_Z\Pi_\lambda)/\Tr(\Pi_\lambda) 
\end{align}
Using a suitable group, every irreducible component has a nonzero coefficient $f_\lambda$ and inverting the measurement channel corresponds to inverting the coefficients in front of the projectors: $\mathcal M^{-1} = \sum_{\lambda\in R_{\mathbb G}}f_\lambda^{-1}\Pi_\lambda$. Note that $\mathcal M^{-1}$ is typically not a physical channel.

So far, we have considered the idealized, noiseless implementation of a randomized measurement twirl.
In practice, however, the  realization of the gates as well as the computational basis measurement will be subject to noise. 
Both read-out noise and gate-independent, non-Markovian gate noise that acts after the gate can be modeled by a map $\Lambda_L$ entering the measurement map as 
\begin{align}
    \label{eq:measurement_channel_M_noisy}
    \widetilde{\mathcal M } = \mathop{\mathbb E}_{U\in\mathbb G}\omega(U)^\dagger\mathcal M_Z\Lambda_L\omega(U) = \sum_{\lambda\in R_{\mathbb G}}\tilde f_\lambda \Pi_\lambda, 
\end{align}
with noisy expansion coefficients
\begin{align}
    \label{eq:expansion_coefficients_noisy}
    \tilde f_\lambda = {\Tr(\mathcal M_Z\Lambda_L\Pi_\lambda)}/{\Tr(\Pi_\lambda)}\,.    
\end{align}

Note that these noise assumptions do not capture general gate-independent noise (acting also before the gate) or gate-dependent noise.
General noise can indeed make $\widetilde {\mathcal M}$ deviate from the form \eqref{eq:measurement_channel_M_noisy}  \cite{Brieger2023CompressiveGST, Brieger2025StabilityShadows}. 

In a nutshell, as long as the noise does not break the symmetry of the original measurement map, the robust shadows protocol reduces to first estimating 
the noisy expansion coefficients, and then using $\tilde{f}_\lambda^{-1}$ instead of $f_\lambda^{-1}$ in the classical post-processing step of the shadow estimation.

In this work, we focus on local Clifford unitaries on $n$ qubits $\mathbb G = {\sf Cl}_2^{\otimes n}$, where each $U\in{\sf Cl}_2^{\otimes n}$ is represented as $U= \bigotimes_{j=1}^n U_j$ for $U_j\in{\sf Cl}_2$.
Twirling over this local group structure imposes a decomposition into $2^n$ irreducible subspaces consisting of combinations of trivial and traceless subspaces for each qubit.
We label them by bit strings $\lambda\in\{0,1\}^n$ and denote the corresponding projector as $\Pi_\lambda = \bigotimes_{k=1}^n\Pi_{\lambda_k}$, where
$\Pi_0 = |\sigma_{0})\!(\sigma_{0}|$ projects onto the identity component, and
$\Pi_1 = \mathbb I_4 - |\sigma_{0})\!(\sigma_{0}|$ projects onto the traceless subspace.
The noisy expansion coefficients Eq.~\eqref{eq:expansion_coefficients_noisy} simplify to
\begin{align}
    \tilde f_\lambda = (\sigma_Z^\lambda | \Lambda_L | \sigma_Z^\lambda)/3^{|\lambda|}\,,
\end{align}
where a linear operator
$A^\lambda = \bigotimes_{j=1}^n A^{\lambda_j}$.

\subsubsection*{Calibration}
Ref.~\cite{chen_robust_2021} proposes the following protocol to estimate $\tilde f_\lambda$:
(i) prepare the trusted ground state $|0 \rangle$ , (ii) apply a random $U\sim\mathbb G$ to $|0 \rangle$, (iii) measure in the $Z$-basis and (iv) store outcome string $b\in\{0,1\}^n$ and gate $U$. 
Repeat this primitive $T_c$ times.
For the acquired data, we then calculate the single-shot estimators for all expansion coefficients, 
which for $\mathbb G = {\sf Cl}_2^{\otimes n}$ reads
$\hat{f}_\lambda = \langle b | Z^\lambda \omega(U)^\dagger | b\rangle$ and 
take the empirical mean $\bar f_\lambda = T^{-1}\sum_{t=1}^T\hat f_{\lambda,t}$. 
Independence ensures that the empirical mean converges to $\tilde f_\lambda$.

\subsubsection*{Shadow estimation} 
The data acquisition of the actual shadow estimation is analogous to the calibration experiment with the difference that the initial state is now the state $\rho$ under scrutiny. 
We now apply random gates on $\rho$ before measuring in the computational basis.  
The acquired shadow data after $T$ iterations are the tuples  $\left\{ (U_t,b_t) \mid t \in \{1, \ldots, T \}\right\}$ of applied unitaries and measurement outcomes. 
To estimate target observables, 
we build robust classical shadow snapshots  $\hat\rho_t = \widetilde{\mathcal{M}}^{-1}(U_t^\dagger|b_t\rangle\!\langle b_t|U_t)$, where $\widetilde{\mathcal{M}}^{-1}$ is calculated from the estimates $\bar f_\lambda$. 
Subsequently, we form an empirical average $\hat{o} = (1/T) \sum_{t=1}^T (O| \hat{\rho}_t)$. 
Under the noise assumptions above and in the limit of infinite samples in the calibration and shadow estimation, $\hat{o}$ converges to 
$(O|\rho)$.
The estimation procedure extends to non-linear properties like state overlap $(\hat\rho|\hat\sigma)$ by empirically averaging over distinct classical shadow snapshots as we detail below.

\subsubsection*{Local noise and snapshots} 
The number of calibration parameters $\tilde f_\lambda$ is exponential in the system size.  
However, if the measurement noise acts locally, i.e.\ in the absence of cross-talk, the parameters factorize into the product of $n$ single-qubit terms $\tilde f_i$. 
For $U=\bigotimes_{j=1}^n U_j$ and outcome string $b\in\{0,1\}^n$ the robust classical snapshot thus becomes
\begin{align}
    \hat\rho = \bigotimes_{j=1}^n\left(\tilde f_j^{-1} U_j^\dagger X_j |b_j\rangle\!\langle b_j|X_jU_j + \frac{1-\tilde f_j^{-1}}{2}\mathbb I \right)\,.
\end{align}
The product structure 
allows us to estimate local observables or overlaps by marginalizing the shadow data to only the relevant qubits. 

\subsection{Robust inner product estimation}
For two states $\rho,\sigma$ measured in bases $U,U'$ with outcomes $b,b'$, respectively, the robust single shot estimator for their overlap is 
\begin{align}
    \label{eq:naive-purity-estimator}
    \Tr(\hat\rho\hat\sigma) = \prod_{i=1}^n \left((\tilde f^{-1}_i)^2\left(\delta_{b_ib_i'}\delta_{U_iU_i'}-\frac{\delta_{U_iU_i'}}{2}\right)+\frac{1}{2}\right)\,.
\end{align}
When setting $\sigma$ to $\rho$, this recovers the purity $\Tr(\rho^2)$.
If the local bases differ ($U_i\neq U_i'$), this term contributes a constant factor $1/2$.
Since this occurs with probability $\mathrm{Pr}(U=U') = 2/3$, the expectation value splits into two cases:
$\mathbb E\left[\Tr(\hat\rho\hat\rho')\right] = 2/3\cdot 1/2 + (1-2/3)\mathbb E\left[\Tr(\hat\rho\hat\rho')|U=U'\right]$. We can avoid this bias by using the same basis $U=U'$ for each state and obtain the compact unbiased estimator
\begin{align}
    \label{eq:brydges-purity-estimator}
    \Tr(\hat\rho\hat\sigma) 
    = \mathop{\mathbb E}
    \left[\prod_{i=1}^n\frac12\left(\frac{(\tilde f^{-1}_i)^2}{3}(-1)^{b_i+b_i'}+1\right)\right].
\end{align}
Setting  $\tilde f^{-1}_i=3$ (no noise) for all $i$ and $\rho = \sigma$, we recover the Rényi entropy estimation protocol of Ref.~\cite{brydges_probing_2019}. Hence, Eq.~\eqref{eq:brydges-purity-estimator} is a robust generalization thereof.

\subsection{Implementation of the randomized measurements} 

In the absence of noise, standard local Clifford shadows can be implemented only with randomized Pauli-bases measurements, requiring a single $\pi/2$-rotation before the read-out. 
Using only $\pi/2$-rotations does, however, not guarantee sufficient twirling of potential noise. 
The original proposal of robust classical shadows 
is thus formulated with uniformly random local Clifford gates for the calibration and the estimation protocol. 
Closer inspection of the structure of the protocol, indicates room for simplifying the implementation that we will explore here.

The single shot estimator Eq.~\eqref{eq:expansion_coefficients_noisy} for local Clifford group expansion coefficients can be written as 
\begin{align}
     \hat f_\lambda 
      &= \frac1{3^{|\lambda|}}\Tr(\bigotimes_{i=1}^n Z^{\lambda_i} U_i^\dagger\frac12\left(\mathbb I + (-1)^{b_i}Z\right)U_i)\,,\nonumber \\
    &= \prod_{i=1}^n\left((-1)^{b_i}\delta_{Z\text{-stab}}(U_i)\right)^{\lambda_i}\,,
\end{align}
where $\delta_{Z\text{-stab}}(U)=1$ if $U^\dagger ZU \in \{\pm Z\}$ and $0$ otherwise.
Only unitaries that stabilize the $Z$-basis contribute to the estimation. 
In the single qubit Clifford group, only $8$ out of $24$ elements satisfy this condition.
Consequently, a large fraction of samples does not contribute to estimating the noisy expansion coefficients of the measurement channel, leading to an inefficient calibration procedure.
This observation motivates restricting the calibration protocol to $Z$-stabilizing operations.
For the all-zero input state this further reduces to applying random bit flips (Pauli-$X$) before measurement. 

Note that this simplified calibration coincides with the protocol of Ref.~\cite{berg_model-free_2022}, where random bit flips are used to characterize read-out noise.
While that work focuses on classical observables, the above derivation shows that the same protocol suffices to estimate the full set of expansion coefficients of the noisy measurement channel. %
This is no coincidence. Calibrating with only the $|0\rangle$ state inevitably requires the assumption of local Clifford-symmetric noise, which in turn can be probed using only a Pauli-$X$ rotation. 

Building on this observation, we can implement the randomized measurement for the shadow estimation using a 
layer of random Pauli-$X$ gates and a layer of Pauli-axis rotations, e.g. a uniform gate from $\{\mathbb I, H, HS^\dagger\}$, where $H$ and $S$ is the Hadamard gate and phase gate, respectively. %
This approach, summarized in Fig.~\ref{fig:main-figure-protocol}, retains the simplicity of the protocol of  Ref.~\cite{berg_model-free_2022} while achieving the generality of robust shadows.

\subsection{Concrete noise model}\label{paragraph:bias}

It is instructive to exemplify the expressions of the robust shadow protocol with a specific, but practically relevant noise model. 
For a single qubit, a classical read-out noise is described by a bit-flip channel. 
For $n$-qubit systems we generally have 
\begin{align}
    \Lambda(\rho) = \sum_{b,b'\in\{0,1\}^n}p_{b\rightarrow b'}|b'\rangle\!\langle b|\rho|b\rangle\!\langle b'|,
\end{align}
such that $\sum_{b'}p_{b\rightarrow b'}=1$ for all bit strings $b$. 
Calculating the expansion coefficients of this error channel yields
\begin{align}
    f_{\lambda} = \frac{(-1)^{|\lambda|}}{3^{|\lambda|}2^n}\sum_{b,b'\in\{0,1\}^n} (-1)^{\langle b, \lambda\rangle + \langle b',\lambda \rangle}p_{b\rightarrow b'},
\end{align}
with inner product $\langle\cdot,\cdot\rangle$.

Because of the Pauli-$X$ symmetry of the twirling, this exactly coincides with expansion coefficients that one would obtain from a stochastic Pauli-$X$ channel 
$\Lambda(\rho) = \sum_{b\in\{0,1\}^n}w_{b}X^b\rho X^b$, where  $f_\lambda = \frac{1}{3^{|\lambda|}}\sum_{b\in\{0,1\}^n}(-1)^{\langle b,\lambda\rangle}w_b$ for $\lambda\in\{0,1\}^n$.

For local bit-flip errors with no crosstalk, the probabilities factorize into products of single qubit probabilities: $p_{b\rightarrow b'}=\prod_{i}p_{b_i\rightarrow b_i'}^{(i)}$ which in turn yields
\begin{align}
    f_i 
    = \frac{1}{3}\left(1 - p_{0\rightarrow1}^{(i)} - p_{1\rightarrow0}^{(i)}\right)
    = \frac{1}{3}\left(1 - 2p_{\mathrm{flip}}^{(i)}\right)\,,
\end{align}
with symmetric bit-flip rate $2p_{\mathrm{flip}}^{(i)}=p_{0\rightarrow1}^{(i)} +  p_{1\rightarrow0}^{(i)}$. 
Note that here we have a one-to-one correspondence between calibration coefficients and symmetric bit-flip rates. 
We can, thus, interpret the calibration coefficients more generally as being determined by \emph{effective symmetric bit-flip rates}.

Uncorrected read-out noise  produces a bias in the standard shadow estimators: 
For a single qubit,  fixed observable $O$ and underlying state $\rho$, the bias, 
$r \coloneqq \left|(O|\rho) - (O|\mathcal M^{-1}\widetilde{\mathcal{M}}|\rho)\right|$, 
 reads  $r=|(1-3\tilde f)\left((O|\rho) - (O| \mathbb{I}_2)/2\right)|$.
For a single-qubit fidelity estimation, assuming perfect state preparation $O = \rho$, this further simplifies to $r=p_{\mathrm{flip}}$. 
For a two-qubit Pauli correlator $P\otimes P'$, the bias evaluates to $|(P\otimes P'|\rho)||1-9\tilde f_1\tilde f_2|$, while the bias for a quadratic estimator like the (sub-system) purity is $r = |(\rho| \sigma) - (\rho|\widetilde{\mathcal M}\mathcal M^{-2}\widetilde{\mathcal M} |\sigma)|$.
For a single qubit this simplifies to
$r= |((\rho|\sigma)-1/2)(1 - 9\tilde f^2)|$ and for the two-qubit case we obtain
\begin{align*}
    r = &\frac{1}{4} +9\tilde f_{1}^2\left(\frac{(\rho_1|\sigma_1)}{2}-\frac14\right)+9\tilde f_{2}^2\left(\frac{(\rho_2|\sigma_2)}{2}-\frac14\right)\\ 
    &+ (9\tilde f_1\tilde f_2)^2\left(\frac14 - \frac{(\rho_1|\sigma_1)+(\rho_2|\sigma_2)}{2} + (\rho|\sigma)\right)\,,
\end{align*}
with $\rho_2=\Tr_1(\rho)$ and $\rho_1=\Tr_2(\rho)$.

While robust shadows remove this bias, they amplify statistical fluctuations.
In particular, the effective sample complexity increases with the inverse of the read-out fidelities, leading to a variance overhead that scales unfavorably with increasing noise strength.
Consistent with Ref.~\cite{chen_robust_2021}, this highlights a fundamental \emph{bias-variance trade-off}: reducing bias via read-out error mitigation comes at the cost of increased variance.

\subsection{Test cases}
To demonstrate the protocol, we prepare three representative quantum states and estimate different observables motivated by applications.

\subsubsection{Haar random product state}
As a first benchmark we implement a $12$-qubit product state generated by local Haar random unitaries $|\psi_{\mathrm{prod}}\rangle = \bigotimes_{k=1}^{12} V_{k} |0\rangle^{\otimes 12}$ with $V_k\overset{\mathrm{iid}}{\sim}\mathrm{\mu_{\mathrm{Haar}}}$ and estimate the single-qubit fidelities.
This setting provides a simple test case, where the effect of read-out errors can be isolated at the single-qubit level.

\subsubsection{QAOA state (1. layer)}

Next, we implement the first layer of a pretrained QAOA state for a weighted max-cut problem~\cite{farhi2014qaoa}.
The underlying graph is defined by Austrian province capitals, with edge weights given by normalized highway distances.
The ansatz consists of alternating cost and mixer Hamiltonian evolutions.
The cost Hamiltonian is given by the MaxCut Hamiltonian, where each edge $(i,j)$ contributes an entangling unitary $\mathrm{exp}\{-i\gamma_{k}Z_{(i,j)}\}$.
The mixer consists of single-qubit rotations $\mathrm{exp}\{-i\beta_k X_l\}$ applied to each qubit $l$.
In this work, we pre-train an $8$-layer circuit and implement the first layer in hardware with $\gamma = 0.29$ and $\beta = 0.56$ and estimate two-qubit subsystem purities $p^{(2)}$.
This allows for probing entanglement properties connected to the diagnosis of weak barren plateaus, as discussed in Ref.~\cite{sack_avoiding_2022}.

\subsubsection{Pauli-correlation encoding}

As a third use case, we demonstrate
using robust shadows to decode 
solutions to combinatorial optimization problems via the
Pauli-correlation encoding (PCE) introduced in Ref.~\cite{Sciorilli_2025}.
The key idea of PCE is to encode binary variables $x_j \in \{-1,+1\}$ into the sign of a $k$-local Pauli expectation value
$ x_j = \mathrm{sgn}\!\left(\langle P_j \rangle\right)$ for $0\leq j\leq m $ with $m=3\binom{n}{k}$.
o all $m$ correlators can be estimated from only three measurement settings, for any choice of $k$.
We use $n = 5$ qubits, which can support up to $m = 30$ Pauli correlators, and we assign 27 variables, one per European Union capital city. Each city is assigned to a unique qubit pair $(i,j)$ and Pauli basis $P \in \{Z, X, Y\}$.
The quantum state $|\psi(\vec{\theta})\rangle$ is prepared using a parametrized brickwork circuit with $L=4$ layers.
Each layer consists of single-qubit rotations (with axes cycling over $X$, $Y$, and $Z$)
followed by entangling gates with odd (even) layers coupling qubit pairs $(0,1),(2,3)$ ($(1,2),(3,4)$).
The circuit parameters $\vec{\theta}$ are optimized classically using Adam ($\eta=0.02$, $500$ steps) to maximize a soft MaxCut objective with decoding $z_k = \tanh(6.25 \langle P_k \rangle)$.
Importantly, all required Pauli expectation values are directly accessible from the shadow data, since random Pauli measurements already sample the corresponding measurement bases.
However, we note that sign changes cannot be captured when using robust estimators of Pauli correlators, and consequently, the decoding of the PCE state does not improve overall.

\subsection{Experiment setup}
The protocol was experimentally demonstrated with the quantum computer Ibex Q1 from AQT~\cite{aqt_2026}. The system runs with 12 fully connected calcium ion qubits, has an average single-qubit gate fidelity over the entire qubit register of \SI{99.97 \pm 0.01}{\percent} and an average two-qubit gate fidelity of \SI{98.48 \pm 0.39}{\percent}. State detection is performed using electron shelving \cite{
dehmelt2012monoion}, where a \SI{1.5}{\milli\second} pulse of light at \SI{397}{\nano\meter}, resonant with the $\mathrm{S}_{1/2}\leftrightarrow \mathrm{P}_{1/2}$  transition, is combined with repumping light on the $\mathrm{P}_{1/2}\leftrightarrow\mathrm{D}_{3/2}$ transition at 866 nm.
As long as the valence electron of the calcium ion is in the ${S}_{1/2}$ state, the laser excitation on the transition causes scattering of millions of photons per second.
If the electron is in the ${D}_{5/2}$ state instead, no light is scattered and the ion remains dark.
The fluorescence emitted by each calcium ion at \SI{397}{\nano\meter} is recorded by an electron-multiplying CCD camera. For each ion, two photon count distributions corresponding to the ``bright'' and ``dark'' state are identified. The duration of the detection pulse is chosen to establish a clear distinction between these two states while simultaneously minimizing any errors that might arise due to unaccounted decay from the $\mathrm{D}_{5/2}$ state. For our usual detection pulse duration the read-out error rate is on the order of \SI{0.5}{\percent} for each ion.

Operating the machine with long read-out pulses of \SI{1.5}{\milli\second} results in read-out errors that are 
very small. However, their correction requires statistics that are significantly beyond the number of shots that the experimental device can produce in a reasonable time span.
We shortened the detection time to \SI{300}{\micro\second} and \SI{150}{\micro\second}, resulting in increased overlap of the ``bright'' and ``dark'' state photon distributions and thus reduced readout fidelity.
We show that robust shadows successfully mitigate the resulting read-out errors.
Since a single shot includes qubit cooling, state initialization, execution, and read-out, reducing the detection time by a factor of 10 decreases the total duration by at most \SI{5}{\percent} on our current set-up.

For all three read-out pulses, we implement our variant (Fig.~\ref{fig:main-figure-protocol}) of the robust shadows protocol using random Pauli-$X$ and Pauli-bases rotation layers. 
To take drift into account, we divide $T$ executions of each state preparation and measurement into 20 batches, interleave them with each other and execute a short calibration phase after/before each new batch.

\begin{figure*}[!t]
    \centering
    \includegraphics[width=0.98\linewidth]{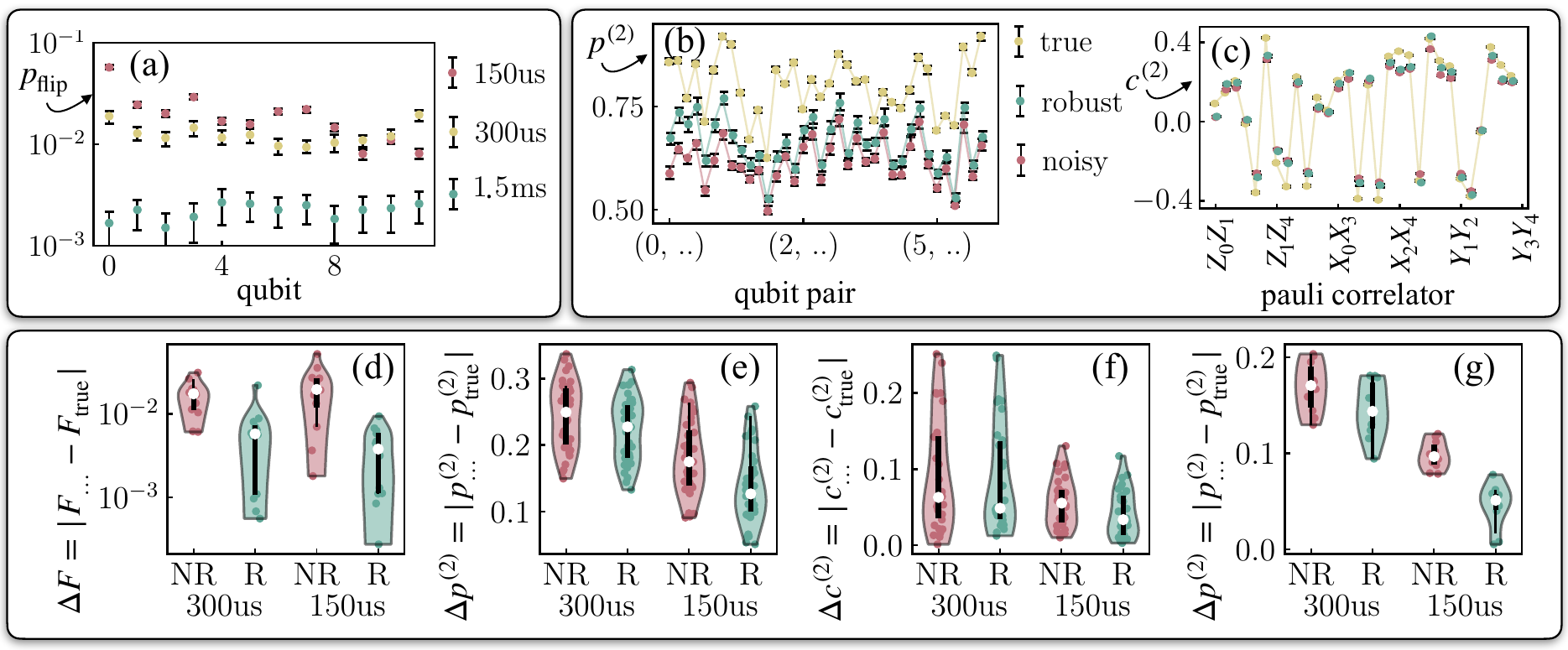}
    \caption{\emph{Experimental data evaluation.}
    Error bars denote 95\% confidence intervals obtained from 20 parametric bootstrap samples.
    (a) Estimated local readout error probabilities $p_{\mathrm{flip}}$ for different measurement pulse lengths, 
    (b) Estimated two-qubit subsystem purities $p^{(2)}$ for the QAOA state across qubit pairs (\SI{150}{\micro\second}).
    (c) Estimated Pauli correlators $c^{(2)} = \langle P_i P_j \rangle$ for the Pauli-correlation encoding (PCE) solution states (\SI{150}{\micro\second}).
    The lower panels 
    show the corresponding absolute deviations $\Delta q = |q - q_{\mathrm{true}}|$ for fidelity of the Haar random product state (d), subsystem purity of the QAOA state (e) and PCE state (f) and their Pauli correlators (g), comparing non-robust (NR) and robust (R) post-processing.
    Left (right) violin plots correspond to a measurement pulse length of \SI{300}{\micro\second} (\SI{150}{\micro\second}).    
    }
    \label{fig:big-main-result}
\end{figure*}

\section{Results} \label{sec:results}

The calibration stage of robust shadows consists of preparing the zero state, applying  random bit flips and  projective read-out.
We then estimate the expansion coefficients from which we can also infer the effective average bit-flip probability per qubit.
In Fig.~\ref{fig:big-main-result} panel (a) the estimated local read-out errors are shown, for \SI{1.5}{\milli\second} ($T_c=12000$) the standard pulse length, \SI{300}{\micro\second} ($T_c=12000$), and \SI{150}{\micro\second} ($T_c=31200$) the shortened pulse lengths.
We observe higher error rates for shorter  measurement pulses as expected. 
The effective bit-flip rates ranges from \SI{0.15}{\percent} to \SI{0.27}{\percent} (mean: \SI{0.22}{\percent}) for the standard pulse, from \SI{0.93}{\percent} to \SI{1.95}{\percent} (mean: \SI{1.28}{\percent}) for the \SI{300}{\micro\second} pulse, and from \SI{0.80}{\percent} to \SI{5.70}{\percent} (mean: \SI{2.06}{\percent}) for the \SI{150}{\micro\second} pulse.

We test the data for cross-talk effects by calculating the non-separability $f_{(0,1)}f_{(1,0)} - f_{(1,1)}$ per qubit pair. 
We find values consistent with zero for all pairs, indicating negligible cross-talk at our statistical precision.

For all test cases, we perform randomized measurements using different read-out pulse lengths and calculate robust shadow estimators using the estimated expansion coefficients. 
The data set with \SI{300}{\micro\second} contains $T=\SI{1e5}{}$ data points whereas the data set \SI{150}{\micro\second} contains $T=\SI{1.6e5}{}$ data points, both where taken on non consecutive different days, the \SI{300}{\micro\second} first, while the \SI{150}{\micro\second} data acquisition had to be split into two non-consecutive different days.

In Fig.~\ref{fig:main-figure-protocol} (right panel) and Fig.~\ref{fig:big-main-result} panels (b.1) and (c.1), we present representative instances of estimates obtained using a \SI{150}{\micro\second} measurement pulse.
Specifically, Fig.~\ref{fig:main-figure-protocol} shows the single-qubit fidelities of the product state, while Fig.~\ref{fig:big-main-result} panel (b.1) displays the two-qubit subsystem purities $p^{(2)}$ of the QAOA state, and panel (c.1) the corresponding Pauli expectation values $c^{(2)} = \langle P_i P_j \rangle$.
In Fig.~\ref{fig:big-main-result} panel (d), (b.2), (c.2) and (c.3) we report the absolute deviations of each quantity  as $\Delta q = | q-q_{\mathrm{true}}|$ for $q$ either the fidelty, subsystem purity, or Pauli expectation value obtained from robust shadows or standard shadows, respectively.
We use the purity estimator Eq.~\eqref{eq:brydges-purity-estimator} for the PCE state and Eq.~\eqref{eq:naive-purity-estimator} for the QAOA state.

In Table~\ref{tab:results} we further summarize and report the average reduction of the absolute deviation when using robust shadows compared to the standard protocol as $\Delta q_{\mathrm{NR}} - \Delta q_{\mathrm{R}}$ and compare them to the theoretical values, see~Sec.~\ref{paragraph:bias}.

Across all observables we observe a consistent reduction of the absolute deviation when applying robust shadows compared to the uncorrected estimator.
This is visible both on the level of individual instances in Fig.~\ref{fig:big-main-result} panels (b) and (c), and in the aggregated statistics shown in panels (d), (e), (f), and (g), where the robust estimates (green) are systematically closer to the true values than the noisy ones (red).
This improvement becomes consistently more pronounced for shorter measurement pulses across all considered quantities.

These observations are in line with the expectation that as the measurement pulse is shortened, the bit-flip probability increases, leading to a larger bias in the uncorrected estimator and consequently a larger achievable improvement through error mitigation.
The quantitative comparison in Table~\ref{tab:results} confirms this behavior. For all observables, the experimentally observed reduction in absolute deviation increases when going from \SI{300}{\micro\second} to \SI{150}{\micro\second}.

While the overall trends are in a good qualitative agreement with the theoretical predictions derived in Sec.~\ref{paragraph:bias}, the experimentally observed reduction in absolute deviation is consistently smaller than predicted.
This can be explained by the fact that theory quantifies bias via expectation values, while experiments estimate finite-sample deviations that are partially masked by statistical fluctuations. Correlations from shared shadow data, finite sampling, and deviations from the noise model further reduce the observed values.

Overall, the experimental results demonstrate that robust shadows reliably reduce readout-induced errors across a range of observables and noise strengths, with performance that is consistent with theoretical expectations up to finite-sample and noise-model effects.

\begin{table}[!t]
    \centering
    \caption{
       Two-qubit bias comparison for different pulse times.}
\renewcommand{\arraystretch}{1.25} 
    \begin{tabular}{l  c c c c}
    \hline
    & \multicolumn{2}{c}{\SI{300}{\micro\second}} & \multicolumn{2}{c}{\SI{150}{\micro\second}} \\
     & Exp. & Theory & Exp. &  Theory \\
    \hline
    (d) $\Delta F_{\mathrm{NR}} - \Delta F_{\mathrm{R}}$ & 0.010(6) & 0.0128(5) & 0.017(4) & 0.0206(5) \\
    (b.2) $\Delta p_{\mathrm{NR}}^{(2)} - \Delta p_{\mathrm{R}}^{(2)}$ & 0.021(2) & 0.037(2) & 0.044(3) & 0.060(1) \\
    (c.2) $\Delta c_{\mathrm{NR}}^{(2)} - \Delta c_{\mathrm{R}}^{(2)}$ & 0.0040(9) & 0.0146(10) & 0.014(2) & 0.0241(7) \\
    (c.3) $\Delta p_{\mathrm{NR}}^{(2)} - \Delta p_{\mathrm{R}}^{(2)}$ & 0.024(4) & 0.035(2) & 0.053(9) & 0.057(1) \\
    \bottomrule
    \end{tabular}
    
    \label{tab:results}
\end{table}

\section{Summary and Outlook}

In this work, we demonstrated a local robust shadows protocol on a trapped-ion quantum computing system and investigated its performance under increased read-out noise.
By shortening the measurement pulse, we increase read-out errors  and characterized their effect on shadow estimators.
We showed that the protocol consistently reduces the resulting bias across a range of observables, including single-qubit fidelities, two-qubit subsystem purities, and Pauli correlators.

However, this bias reduction comes at the cost of an increased estimator variance. 
In our setting, the additional sampling overhead required for calibration and error mitigation outweighs the time saved by shortening the measurement duration on the trapped-ion system. As a result, we do not observe a net gain in efficiency when combining shorter measurement pulses with robust shadow techniques.

Looking forward, several directions remain to be explored. First, it would be interesting to extend our systematic analysis of robust shadows under controlled-noise effects beyond read-out errors to other sources of errors, such as amplitude damping, coherent unitary errors, and stochastic unitary noise.  
Theory suggests that especially coherent, gate-dependent noise potentially affects the performance of robust shadows significantly \cite{Brieger2025StabilityShadows}. 
Second, a more systematic investigation of the bias-variance trade-off as a function of the measurement pulse length could clarify whether there exist regimes in which shorter measurements combined with mitigation lead to an overall advantage.
Finally, the calibration data obtained from our robust shadow protocol provides a rich source of information about the device. This data could be leveraged to study temporal drift, identify correlated errors, or fit structured models, such as tensor networks, to better understand and quantify cross-talk and other non-local-noise effects in the system.

\section*{Acknowledgments}
This work was funded by the Austrian Federal Ministry of Education, Science and Research via the Austrian Research Promotion Agency (FFG) under the flagship project HPQC (FO999897481) and the COMET module project QAE (923923). We thank all our colleagues from these consortia, whose support helped advance this research.
We gratefully acknowledge support by the European Union’s Horizon Europe research and innovation program under Grant Agreement Number 101114305 (MILLENION-SGA1), 101113690 (PASQUANS2.1) and by the aws (Austrian Promotional Bank) with funds from the National Foundation for Research, Technology and Development (Fonds Zukunft Österreich) under project nr. P2508223-Q2M01.

\bibliographystyle{myIEEEtran.bst}
\bibliography{mybib} 

@article{Brieger2025StabilityShadows,
  title   = {Stability of Classical Shadows under Gate-Dependent Noise},
  author  = {Brieger, Raphael and Heinrich, Markus and Roth, Ingo and Kliesch, Martin},
  journal = {Physical Review Letters},
  volume  = {134},
  number  = {9},
  pages   = {090801},
  year    = {2025},
  doi     = {10.1103/PhysRevLett.134.090801}
}

@article{Maciejewski2020mitigationofreadout,
  doi = {10.22331/q-2020-04-24-257},
  url = {https://doi.org/10.22331/q-2020-04-24-257},
  title = {Mitigation of readout noise in near-term quantum devices by classical post-processing based on detector tomography},
  author = {Maciejewski, Filip B. and Zimbor{\'{a}}s, Zolt{\'{a}}n and Oszmaniec, Micha{\l{}}},
  journal = {{Quantum}},
  issn = {2521-327X},
  publisher = {{Verein zur F{\"{o}}rderung des Open Access Publizierens in den Quantenwissenschaften}},
  volume = {4},
  pages = {257},
  month = apr,
  year = {2020}
}

@article{Brieger2023CompressiveGST,
  title   = {Compressive Gate Set Tomography},
  author  = {Brieger, Raphael and Roth, Ingo and Kliesch, Martin},
  journal = {PRX Quantum},
  volume  = {4},
  number  = {1},
  pages   = {010325},
  year    = {2023},
  doi     = {10.1103/PRXQuantum.4.010325}
}

@article{huang_predicting_2020,
	title = {Predicting {Many} {Properties} of a {Quantum} {System} from {Very} {Few} {Measurements}},
	volume = {16},
	issn = {1745-2473, 1745-2481},
	url = {http://arxiv.org/abs/2002.08953},
	doi = {10.1038/s41567-020-0932-7},
	number = {10},
	urldate = {2025-04-22},
	journal = {Nature Physics},
	author = {Huang, Hsin-Yuan and Kueng, Richard and Preskill, John},
	month = oct,
	year = {2020},
	note = {arXiv:2002.08953 [quant-ph]},
	pages = {1050--1057},
}

@article{berg_model-free_2022,
	title = {Model-free readout-error mitigation for quantum expectation values},
	volume = {105},
	issn = {2469-9926, 2469-9934},
	url = {http://arxiv.org/abs/2012.09738},
	doi = {10.1103/PhysRevA.105.032620},
	number = {3},
	urldate = {2025-01-08},
	journal = {Physical Review A},
	author = {Berg, Ewout van den and Minev, Zlatko K. and Temme, Kristan},
	month = mar,
	year = {2022},
	note = {arXiv:2012.09738 [quant-ph]},
	pages = {032620},
}

@article{chen_robust_2021,
	title = {Robust shadow estimation},
	volume = {2},
	issn = {2691-3399},
	url = {http://arxiv.org/abs/2011.09636},
	doi = {10.1103/PRXQuantum.2.030348},
	number = {3},
	urldate = {2025-01-08},
	journal = {PRX Quantum},
	author = {Chen, Senrui and Yu, Wenjun and Zeng, Pei and Flammia, Steven T.},
	month = sep,
	year = {2021},
	note = {arXiv:2011.09636 [quant-ph]},
	pages = {030348},
}

@article{Farias_2025,
   title={Robust ultra-shallow shadows},
   volume={10},
   ISSN={2058-9565},
   url={http://dx.doi.org/10.1088/2058-9565/adc14f},
   DOI={10.1088/2058-9565/adc14f},
   number={2},
   journal={Quantum Science and Technology},
   publisher={IOP Publishing},
   author={Farias, Renato M S and Peddinti, Raghavendra D and Roth, Ingo and Aolita, Leandro},
   year={2025},
   month=mar, pages={025044} }

@article{brydges_probing_2019,
	title = {Probing {Rényi} entanglement entropy via randomized measurements},
	volume = {364},
	url = {https://www.science.org/doi/full/10.1126/science.aau4963},
	doi = {10.1126/science.aau4963},
	number = {6437},
	urldate = {2024-03-25},
	journal = {Science},
	author = {Brydges, Tiff and Elben, Andreas and Jurcevic, Petar and Vermersch, Benoît and Maier, Christine and Lanyon, Ben P. and Zoller, Peter and Blatt, Rainer and Roos, Christian F.},
	month = apr,
	year = {2019},
	note = {Publisher: American Association for the Advancement of Science},
	pages = {260--263},
}

@article{sack_avoiding_2022,
	title = {Avoiding barren plateaus using classical shadows},
	volume = {3},
	issn = {2691-3399},
	url = {http://arxiv.org/abs/2201.08194},
	doi = {10.1103/PRXQuantum.3.020365},
	number = {2},
	urldate = {2025-12-09},
	journal = {PRX Quantum},
	author = {Sack, Stefan H. and Medina, Raimel A. and Michailidis, Alexios A. and Kueng, Richard and Serbyn, Maksym},
	month = jun,
	year = {2022},
	note = {arXiv:2201.08194 [quant-ph]},
	pages = {020365},
}

@article{hu_demonstration_2025,
	title = {Demonstration of {Robust} and {Efficient} {Quantum} {Property} {Learning} with {Shallow} {Shadows}},
	volume = {16},
	issn = {2041-1723},
	url = {http://arxiv.org/abs/2402.17911},
	doi = {10.1038/s41467-025-57349-w},
	number = {1},
	urldate = {2025-12-15},
	journal = {Nature Communications},
	author = {Hu, Hong-Ye and Gu, Andi and Majumder, Swarnadeep and Ren, Hang and Zhang, Yipei and Wang, Derek S. and You, Yi-Zhuang and Minev, Zlatko and Yelin, Susanne F. and Seif, Alireza},
	month = mar,
	year = {2025},
	note = {arXiv:2402.17911 [quant-ph]},
	pages = {2943},
}

@article{joshi_mpemba_2024,
  title = {Observing the Quantum Mpemba Effect in Quantum Simulations},
  author = {Joshi, Lata Kh. and Franke, Johannes and Rath, Aniket and Ares, Filiberto and Murciano, Sara and Kranzl, Florian and Blatt, Rainer and Zoller, Peter and Vermersch, Beno\^{\i}t and Calabrese, Pasquale and Roos, Christian F. and Joshi, Manoj K.},
  journal = {Phys. Rev. Lett.},
  volume = {133},
  issue = {1},
  pages = {010402},
  numpages = {7},
  year = {2024},
  month = {7},
  publisher = {American Physical Society},
  doi = {10.1103/PhysRevLett.133.010402},
  url = {https://link.aps.org/doi/10.1103/PhysRevLett.133.010402}
}

@article{joshi_fcs_2025,
   title={Measuring Full Counting Statistics in a Trapped-Ion Quantum Simulator},
   volume={135},
   ISSN={1079-7114},
   url={http://dx.doi.org/10.1103/gyvf-s5bd},
   DOI={10.1103/gyvf-s5bd},
   number={16},
   journal={Physical Review Letters},
   publisher={American Physical Society (APS)},
   author={Joshi, Lata Kh and Ares, Filiberto and Joshi, Manoj K. and Roos, Christian F. and Calabrese, Pasquale},
   year={2025},
   month=oct }

@article{Sciorilli_2025,
   title={Towards large-scale quantum optimization solvers with few qubits},
   volume={16},
   ISSN={2041-1723},
   url={http://dx.doi.org/10.1038/s41467-024-55346-z},
   DOI={10.1038/s41467-024-55346-z},
   number={1},
   journal={Nature Communications},
   publisher={Springer Science and Business Media LLC},
   author={Sciorilli, Marco and Borges, Lucas and Patti, Taylor L. and García-Martín, Diego and Camilo, Giancarlo and Anandkumar, Anima and Aolita, Leandro},
   year={2025},
   month=jan }

@article{moses_race_2023,
	title = {A {Race} {Track} {Trapped}-{Ion} {Quantum} {Processor}},
	volume = {13},
	issn = {2160-3308},
	url = {http://arxiv.org/abs/2305.03828},
	doi = {10.1103/PhysRevX.13.041052},
	number = {4},
	urldate = {2026-03-18},
	journal = {Physical Review X},
	author = {Moses, S. A. and Baldwin, C. H. and Allman, M. S. and Ancona, R. and Ascarrunz, L. and Barnes, C. and Bartolotta, J. and Bjork, B. and Blanchard, P. and Bohn, M. and Bohnet, J. G. and Brown, N. C. and Burdick, N. Q. and Burton, W. C. and Campbell, S. L. and Campora, J. P. and Carron, C. and Chambers, J. and Chan, J. W. and Chen, Y. H. and Chernoguzov, A. and Chertkov, E. and Colina, J. and Curtis, J. P. and Daniel, R. and DeCross, M. and Deen, D. and Delaney, C. and Dreiling, J. M. and Ertsgaard, C. T. and Esposito, J. and Estey, B. and Fabrikant, M. and Figgatt, C. and Foltz, C. and Foss-Feig, M. and Francois, D. and Gaebler, J. P. and Gatterman, T. M. and Gilbreth, C. N. and Giles, J. and Glynn, E. and Hall, A. and Hankin, A. M. and Hansen, A. and Hayes, D. and Higashi, B. and Hoffman, I. M. and Horning, B. and Hout, J. J. and Jacobs, R. and Johansen, J. and Jones, L. and Karcz, J. and Klein, T. and Lauria, P. and Lee, P. and Liefer, D. and Lytle, C. and Lu, S. T. and Lucchetti, D. and Malm, A. and Matheny, M. and Mathewson, B. and Mayer, K. and Miller, D. B. and Mills, M. and Neyenhuis, B. and Nugent, L. and Olson, S. and Parks, J. and Price, G. N. and Price, Z. and Pugh, M. and Ransford, A. and Reed, A. P. and Roman, C. and Rowe, M. and Ryan-Anderson, C. and Sanders, S. and Sedlacek, J. and Shevchuk, P. and Siegfried, P. and Skripka, T. and Spaun, B. and Sprenkle, R. T. and Stutz, R. P. and Swallows, M. and Tobey, R. I. and Tran, A. and Tran, T. and Vogt, E. and Volin, C. and Walker, J. and Zolot, A. M. and Pino, J. M.},
	month = dec,
	year = {2023},
	note = {arXiv:2305.03828 [quant-ph]},
	pages = {041052},
}

@article{campbell_path_2025,
	title = {A {Path} to {Scalable} {Quantum} {Computers}},
	volume = {18},
	copyright = {©2025 by the American Physical Society. All rights reserved.},
	url = {https://physics.aps.org/articles/v18/40},
	doi = {10.1103/PhysRevX.15.011040},
	language = {en},
	urldate = {2026-03-18},
	journal = {Physics},
	publisher = {American Physical Society},
	author = {Campbell, Sara},
	month = feb,
	year = {2025},
	pages = {40}
}

@article{Lowe_2025,
   title={Lower Bounds for Learning Quantum States with Single-Copy Measurements},
   volume={17},
   ISSN={1942-3462},
   url={http://dx.doi.org/10.1145/3717450},
   DOI={10.1145/3717450},
   number={1},
   journal={ACM Transactions on Computation Theory},
   publisher={Association for Computing Machinery (ACM)},
   author={Lowe, Angus and Nayak, Ashwin},
   year={2025},
   month=mar, pages={1–42} }

@article{Zhao_2024,
   title={Group-theoretic error mitigation enabled by classical shadows and symmetries},
   volume={10},
   ISSN={2056-6387},
   url={http://dx.doi.org/10.1038/s41534-024-00854-5},
   DOI={10.1038/s41534-024-00854-5},
   number={1},
   journal={npj Quantum Information},
   publisher={Springer Science and Business Media LLC},
   author={Zhao, Andrew and Miyake, Akimasa},
   year={2024},
   month=jun }

@article{Elben_2022,
   title={The randomized measurement toolbox},
   volume={5},
   ISSN={2522-5820},
   url={http://dx.doi.org/10.1038/s42254-022-00535-2},
   DOI={10.1038/s42254-022-00535-2},
   number={1},
   journal={Nature Reviews Physics},
   publisher={Springer Science and Business Media LLC},
   author={Elben, Andreas and Flammia, Steven T. and Huang, Hsin-Yuan and Kueng, Richard and Preskill, John and Vermersch, Benoît and Zoller, Peter},
   year={2022},
   month=dec, pages={9–24} }

@misc{aqt_2026,
  author = {{Alpine Quantum Technologies GmbH}},
  title = {{Ibex Q1 Quantum Computer System}},
  year = {2026},
  url = {https://www.aqt.eu/products/ibex-q1/},
  note = {Accessed: 2025-03-24}
}

@article{Hainzer_2024,
  title = {Correlation Spectroscopy with Multiqubit-Enhanced Phase Estimation},
  author = {Hainzer, H. and Kiesenhofer, D. and Ollikainen, T. and Bock, M. and Kranzl, F. and Joshi, M. K. and Yoeli, G. and Blatt, R. and Gefen, T. and Roos, C. F.},
  journal = {Phys. Rev. X},
  volume = {14},
  issue = {1},
  pages = {011033},
  numpages = {24},
  year = {2024},
  month = {02},
  publisher = {American Physical Society},
  doi = {10.1103/PhysRevX.14.011033},
  url = {https://link.aps.org/doi/10.1103/PhysRevX.14.011033}
}

@article{Kiesenhofer_2023,
  title = {Controlling Two-Dimensional Coulomb Crystals of More Than 100 Ions in a Monolithic Radio-Frequency Trap},
  author = {Kiesenhofer, Dominik and Hainzer, Helene and Zhdanov, Artem and Holz, Philip C. and Bock, Matthias and Ollikainen, Tuomas and Roos, Christian F.},
  journal = {PRX Quantum},
  volume = {4},
  issue = {2},
  pages = {020317},
  numpages = {16},
  year = {2023},
  month = {4},
  publisher = {American Physical Society},
  doi = {10.1103/PRXQuantum.4.020317},
  url = {https://link.aps.org/doi/10.1103/PRXQuantum.4.020317}
}

@article{Kranzl_2022,
  title = {Controlling long ion strings for quantum simulation and precision measurements},
  author = {Kranzl, Florian and Joshi, Manoj K. and Maier, Christine and Brydges, Tiff and Franke, Johannes and Blatt, Rainer and Roos, Christian F.},
  journal = {Phys. Rev. A},
  volume = {105},
  issue = {5},
  pages = {052426},
  numpages = {13},
  year = {2022},
  month = {5},
  publisher = {American Physical Society},
  doi = {10.1103/PhysRevA.105.052426},
  url = {https://link.aps.org/doi/10.1103/PhysRevA.105.052426}
}

@article{dehmelt2012monoion,
  title={Monoion oscillator as potential ultimate laser frequency standard},
  author={Dehmelt, Hans G},
  journal={IEEE transactions on instrumentation and measurement},
  number={2},
  pages={83--87},
  year={2012},
  publisher={IEEE}
}

@article{Banaszek_2013,
doi = {10.1088/1367-2630/15/12/125020},
url = {https://doi.org/10.1088/1367-2630/15/12/125020},
year = {2013},
month = {12},
publisher = {IOP Publishing},
volume = {15},
number = {12},
pages = {125020},
author = {Banaszek, K and Cramer, M and Gross, D},
title = {Focus on quantum tomography},
journal = {New Journal of Physics},
}

@inproceedings{Haah2016,
author = {Haah, Jeongwan and Harrow, Aram W. and Ji, Zhengfeng and Wu, Xiaodi and Yu, Nengkun},
title = {Sample-optimal tomography of quantum states},
year = {2016},
isbn = {9781450341325},
publisher = {Association for Computing Machinery},
address = {New York, NY, USA},
url = {https://doi.org/10.1145/2897518.2897585},
doi = {10.1145/2897518.2897585},
booktitle = {Proceedings of the Forty-Eighth Annual ACM Symposium on Theory of Computing},
pages = {913–925},
numpages = {13},
location = {Cambridge, MA, USA},
series = {STOC '16}
}

@article{Kueng2017,
title = {Low rank matrix recovery from rank one measurements},
journal = {Applied and Computational Harmonic Analysis},
volume = {42},
number = {1},
pages = {88-116},
year = {2017},
issn = {1063-5203},
doi = {https://doi.org/10.1016/j.acha.2015.07.007},
url = {https://www.sciencedirect.com/science/article/pii/S1063520315001037},
author = {Richard Kueng and Holger Rauhut and Ulrich Terstiege}
}

@misc{farhi2014qaoa,
      title={A Quantum Approximate Optimization Algorithm}, 
      author={Edward Farhi and Jeffrey Goldstone and Sam Gutmann},
      year={2014},
      eprint={1411.4028},
      archivePrefix={arXiv},
      primaryClass={quant-ph},
      url={https://arxiv.org/abs/1411.4028}, 
}

@article{vitale_2024,
   title={Robust Estimation of the Quantum Fisher Information on a Quantum Processor},
   volume={5},
   ISSN={2691-3399},
   url={http://dx.doi.org/10.1103/PRXQuantum.5.030338},
   DOI={10.1103/prxquantum.5.030338},
   number={3},
   journal={PRX Quantum},
   publisher={American Physical Society (APS)},
   author={Vitale, Vittorio and Rath, Aniket and Jurcevic, Petar and Elben, Andreas and Branciard, Cyril and Vermersch, Benoît},
   year={2024},
   month=aug }

\end{document}